\documentclass{raa}
\usepackage{amsfonts}
\usepackage{mathrsfs}          % referee version: for submission
\usepackage{natbib}
\usepackage{graphicx,times}             %for PS/EPS graphics inclusion, new

\newcommand{\Mpc}{\ensuremath{{\rm Mpc}}}

\begin{document}
\title{Primordial Non-Gaussianity from LAMOST Surveys}

\volnopage{Vol.0 (200x) No.0, 000--000}      %%preserved for Editor. DOn't remove!
\setcounter{page}{1}           %%starting page, preserved for Editor. DOn't remove!

\author{Yan Gong\inst{1,2}
\and
        Xin Wang\inst{1,2}
\and
        Zheng Zheng\inst{3}
\and
        Xuelei Chen\inst{1,4,5}\mailto{}
      }

\institute{$^1$National Astronomical Observatories, Chinese Academy of Sciences,
Beijing 100012, China\\
$^2$Graduate School of Chinese Academy of Sciences, Beijing 100049, China\\
$^3$Institute for Advanced Study, Einstein Drive, Princeton, NJ 08540, USA\\
$^4$Center of High Energy Physics, Peking University, Beijing 100871, China\\
$^5$Kavli Institute for Theoretical Physics China, CAS, Beijing 100190, China\\
\email{xuelei@cosmology.bao.ac.cn}}

\date{Received~~\today; accepted~~}

\abstract{The primordial non-Gaussianity (PNG) in matter density perturbation
 is a very powerful probe of the physics of the very early Universe. The local
PNG can induce a distinct scale-dependent bias on the large scale
structure distribution of galaxies and quasars, which could be used
for constraining it. We study the detection limits on PNG from the
surveys of the LAMOST telescope. The cases of the main galaxy
survey, the luminous red galaxy (LRG) survey, and the quasar survey
of different magnitude limits are considered. We find that the MAIN1
sample (i.e. the main galaxy survey with one magnitude deeper than
the SDSS main galaxy survey, or $r<18.8$) could only provide very
weak constraint on PNG. For the MAIN2 sample ($r<19.8$) and the LRG
survey, the $2\sigma (95.5\%)$ limit on the PNG parameter $f_{\rm
NL}$ are $|f_{\rm NL}|<145$ and $|f_{\rm NL}|<114$ respectively,
comparable to the current limit from cosmic microwave background
(CMB) data. The quasar survey could provide much more stringent
constraint, and we find that the $2\sigma$ limit for $|f_{\rm NL}|$ is
between 50 and 103, depending on the magnitude limit of the survey.
With Planck-like priors on cosmological parameters, the quasar
survey with $g<21.65$ would improve the constraints to $|f_{\rm
NL}|<43$ ($2\sigma$). We also discuss the possibility of further
tightening the constraint by using the relative bias method proposed
by \cite{selj08}. 
 \\
{\tt preprint number: CAS-KITPC/ITP-161, arxiv:0904.4257}
}

%   \authorrunning{ }            %author_head in even pages
   \titlerunning{Primordial Non-Gaussianity from the LAMOST
   Surveys}  % title_head in odd pages

   \maketitle

\section{Introduction}
\label{sect:intro}
The seed of large scale structure (LSS) is usually thought to be produced
during a period of inflation in the early universe. Models of the
standard scenario of single field slow roll inflation generally predict that
these primordial perturbations are Gaussian distributed
\citep{mald03,acqu03,crem03,bart04}. However, significant primordial
non-Gaussianity (PNG) may be produced by multi-field inflation \citep{lind97}
and reheating \citep{dvali04}, if the inflaton potential has some sharp
features \citep{CEL07},
or if the inflation is driven by field with non-canonical kinetic
terms \citep{CHKS07}.
PNG is also predicted by alternative models to
inflation, such as the ekpyrotic scenario \citep{lehn08}. Thus,
non-Gaussianity in primordial density perturbations could be a a very powerful
discriminator for the physical models of the very early
Universe \citep{bart04,koma09}.

The PNG in density perturbation is usually parametrized as
 \be
 \Phi({\bf x})=\phi({\bf x})+f_{\rm NL}(\phi^2({\bf x})-\langle\phi^2\rangle),
 \ee
where $\Phi$ is Bardeen's gauge-invariant potential and $\phi$ is a
Gaussian random field. The parameter $f_{\rm NL}$, which denotes the
effect of the PNG, is typically of order $10^{-2}$ in standard inflation
models (e.g., \citealt{Gangui94}). Non-linear transformation from primordial
field fluctuation to observables generically gives rise to
$f_{\rm NL} \sim O(1)$. With $\phi \sim 10^{-5}$, this produces an extremely
small non-Gaussianity in observables, such as the cosmic microwave background
(CMB) temperature anisotropy or LSS bispectrum.

The CMB anisotropy so far offers the cleanest
observational probe of PNG \citep{wang00,verd01}.
\cite{yada08} recently claimed
detection of a positive $f_{\rm NL}$ at 99.5\%
significance ($27<f_{\rm NL}<147$ at $2\sigma$) from
the Wilkinson Microwave Anisotropy Probe (WMAP) 3-year data.
Analysis of the WMAP 5-year data using the bispectrum
also favors a slightly positive $f_{\rm NL}$,
$-9<f_{\rm NL}<111$ at $95.5\%$ confidence level (C.L.) \citep{koma08}, while
the Minkowski functional method yields  $-178<f_{\rm NL}<64$ \citep{koma08}.
In the future, the Planck satellite should
significantly improve the constraint on $f_{\rm NL}$, which may eventually push
the uncertainty in $f_{\rm NL}$ to be around 5 \citep{koma01}.
Alternatively, one may use the LSS observations to search for the
PNG. However, even if the initial fluctuations are
Gaussian, during the gravitational growth
of the LSS, non-linear effect also generates non-Gaussianity, which
may swamp the primordial one. Typically,
the LSS constraints on PNG using the bispectrum technique
are weak \citep{verd01,scoc04,sefu07}.
One consequence of PNG in structure formation is to significantly enhance
the number of the rare peaks of fluctuation, boosting the formation of
massive objects \citep{lucc86,mata00}. One could
observe this effect from the abundance of galaxy clusters \citep{bens02} or
dark matter halos harboring the first generation of stars \citep{chen03}.
However, the required observational task is difficult due to low-number
statistics.

Recently, \cite{dala08} proposed a powerful and practical method to
probe the local PNG with LSS observations. The PNG-caused enhancement on the
formation of massive dark matter halos induces a distinctive scale-dependent
bias on the largest scales. Subsequent works confirmed and extended this
result \citep{mata08,slos08,afsh08,mcdo08,carb08,kami08}, showing that useful limits
on PNG could be obtained from LSS surveys.

In the present work, we investigate the sensitivity to PNG from the
extragalactic surveys of the Large sky Area Multi-Object Spectroscopic Telescope
(LAMOST)\footnote{{\it http://www.lamost.org/}}.
The LAMOST is a recently built Chinese 4-meter
Schmidt telescope with a field of view of 20 $\deg^2$.
Equipped with 4000 optical fibers that are individually positioned
by computer during observation, it can take spectra of 4000 targets
simultaneously \citep{C98,S08}.
The LAMOST is currently undergoing engineering tests and calibrations, and it is expected
to begin conducting surveys by 2010/2011.
In a recent paper, we have forecasted the constraining power of the LAMOST extragalactic
surveys on dark energy measurement \citep{wang08} by considering different samples of
galaxies and quasars. Here, we make similar forecast on the constraining power on PNG.

\section{Method}
\label{sect:Obs}
On large scales, the PNG induces a distinct scale-dependent bias in the
distribution of tracer objects, which can be measured
from the correlation function or power spectrum of galaxies and quasars. Extending the work
of \cite{dala08}, \cite{slos08} found that this bias departures from the Gaussian case as
 \be
 \Delta b(z,M,k)=2f_{\rm NL}(b-p)\delta_c\frac{3\Omega_{m0}}{2k^2T(k)D(z)}\Big(\frac{H_0}{c}\Big)^2,
 \ee
where $b$ is the Eulerian bias and can be calculated as usual  with the
halo model, $\delta_c=1.686$, $D(z)$ is the linear growth factor
normalized to be the scale factor $a$ in the matter domination epoch,
$\Omega_{m0}$
is the present matter energy density parameter, $T(k)$ is the
transfer function, $H_0$ is the current Hubble parameter and $c$ is
the speed of light.

The parameter $p$ is introduced to account for the effect of merger bias.
If the effect of merger on the formation of tracer objects is negligible,
its value should be unity. However, the formation of some
objects may be trigger by mergers. For example, quasars is believed to be
accreting massive black holes in the center of galaxies, and the activity of
quasar may be triggered by galaxy mergers, which send gas to the center
of the galaxy, providing the materials needed for accretion.
If the merger effect is dominant, $p=1.6$. In general, the value of $p$
is between 1 and 1.6.

 We consider three types of extragalactic surveys for LAMOST.
(1) The main galaxy survey, which is a magnitude-limited general survey
of all types of galaxies. By MAIN1 and MAIN2, we denote respectively the samples
which are 1 and 2 magnitude deeper than the SDSS surveys, i.e. MAIN1 for
$r<18.8$, and MAIN2 for $r<19.8$. (2) The LRG survey, a survey of
color-selected luminous red galaxies (LRGs). Here for illustration
we have used the MegaZ LRG sample \citep{coll06} for calculation,
though in reality the LAMOST LRG will be selected independently. (3) The
quasar survey. Depending on the magnitude limit, we define
three samples, QSO1 ($g<20.5$), QSO2 ($g<21$), and QSO3 ($g<21.65$).
As in \cite{wang08}, we assume that the
surveyed area is $8000\deg^2$, for which we have the SDSS imaging data,
even though the actual survey may cover a larger area if input catalogue
for the south Galactic cap is available. For more details of these samples,
including their number densities, redshift distributions, and the
values of bias (estimated in the Gaussian models), see \cite{wang08}.

We model the galaxy power spectrum as
 \be
 P_g(k)=(b_g+\Delta b)^2P_{\rm lin}(k)\frac{1+Qk^2}{1+Ak},
 \ee
where $b_g$is the bias of the galaxies calculated for the Gaussian case,
$P_{\rm lin}$ is the linear
matter power spectrum \citep{eise98}, $k$ is the wavenumber
($10^{-2}\ h{\rm Mpc}^{-1} < k < 0.2\ h {\rm Mpc}^{-1}$),
$Q$ and $A$ are free parameters
which describe the non-linear effect. For fiducial models, we fix $Q=4.6$
for the main samples, $Q=30$ for the LRG sample, and adopt
$A=1.4$ \citep{cole05,tegm06}. For the quasar power spectrum, we adopt the
simpler form
 \be
 P_q(k)=(b_q+\Delta b)^2P_{\rm lin}(k),
 \ee
where $b_q$ is the bias of quasars. The value of bias for each of the sample
has been given in \cite{wang08}. We assume that the PNG is small, so that
the value of bias does not change much.

The deviation of bias could be discovered with precision measurement of
the power spectrum of
the galaxy or quasar distribution. The measurement error of the power spectrum
is given by \citep{feld94, tegm97}
 \be
 \frac{\sigma_P}{P}=2\pi\sqrt{\frac{1}{V_{\rm s}k^2\Delta
 k}}\frac{1+nP}{nP}=2\pi\sqrt{\frac{1}{V_{\rm eff}k^2\Delta k}},
 \ee
where $V_{\rm s}$ is the survey volume and $V_{\rm eff}$ is the
effective volume, defined by
 \be
 V_{\rm eff}\equiv\int\bigg[\frac{n({\bf r})P}{n({\bf r})P+1}\bigg]^2d{\bf r}
= \bigg[\frac{nP}{nP+1}\bigg]^2V_{\rm s}.
 \ee
The second equality holds if the comoving number density $n({\bf r})$ is
a constant. We divide the samples into different redshift bins as
in \cite{wang08},
then the total likelihood function
is the product of the likelihood function for all the different
bins.

\begin{figure}
\begin{center}
\includegraphics[width=0.5\textwidth]{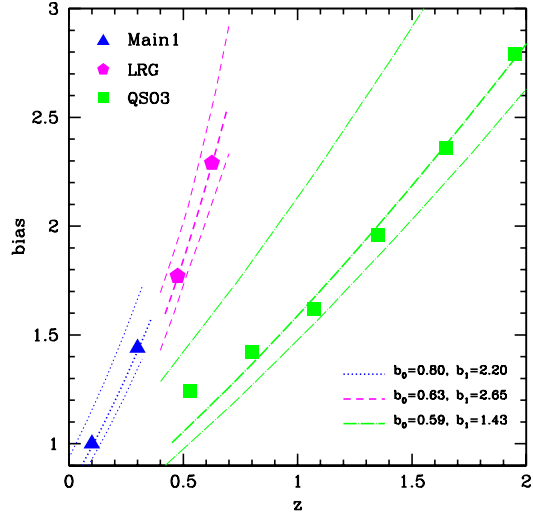}
\caption{\label{fig:b_f} Fiducial values of bias and the fits with the
functional form $b(z)=b_0(1+z)^\gamma$ for the Main1, LRG, and QSO3 samples.
For each sample, the fiducial bias values (symbols) are from \citet{wang08},
the thick curve are the best fit with the thin curves denotes the central
68\% distribution maginalized over the MCMC results.
}
   \end{center}
\end{figure}

In order to estimate the error on the measurement of $f_{\rm NL}$,
we use a Markov Chain Monte Carlo (MCMC) technique to explore the
parameters space, then marginalize over (i.e., integrate out) the
other parameters to obtain a constraint on the PNG. For details of
the MCMC technique, see, e.g., our earlier paper \citep{gong07}. The
fiducial cosmology we adopt is the WMAP 5-year best fit $\Lambda$CDM
model without PNG, with $\Omega_{b0}=0.05$, $\Omega_{m0}=0.24$,
$\Omega_{\Lambda0}=0.76$, $\sigma_8=0.82$, $n_s=0.96$, $f_{\rm
NL}=0.0$ and $h_0=0.7$. The allowed ranges of parameters in the
Monte Carlo are $\Omega_{b0}\in(0,0.1)$, $\Omega_{m0}\in(0,1)$,
$\sigma_8\in(0.6,0.9)$, $n_s\in(0.9,1.1)$, $f_{\rm
NL}\in(-400,400)$, $Q\in(0,40)$ and $h_0\in(0.5,0.9)$.

The value of bias for the sample is an important parameter which
is partially degenerate with $f_{\rm NL}$. We model the
value of bias at different redshifts bins in two possible forms:
\begin{equation}
b(z)=b_0(1+z)^\gamma,
\label{eq:bform1}
\end{equation}
or
\begin{equation}
b(z) = c_1 + c_2/D_0(z)
\label{eq:bform2}
\end{equation}
where $D_0(z)$ is the
growth factor normalized such that $D_0(z=0) = 1$.
%This latter form is inspired by
%$b(z) = 1 + (b_p-1)/D_0(z)$, which can be derived under the assumption
%$\dot{\delta_g}=\dot{\delta_m}$, where $b_p$ is the present day galaxy
%bias, $\delta_g$ and $\delta_m$ are the galaxy and matter density contrast
%respectively.
This latter form is inspired by $b(z)=1+(b_p-1)/D0(z)$, which can be derived
in the linear regime for a passively evolved tracer population without any
mergers by using $\dot{\delta_g}=\dot{\delta_m}$ and
$\delta_g(z)=b(z)\delta_m(z)$ \citep{Fry96}. Here $b_p$ is the present day
bias factor of the tracer, $\delta_g$ and $\delta_m$ are the tracer and
matter density contrast, respectively.
We then try to use these two parametrization forms to fit
the bias for our samples which were given in \citet{wang08}. Our test
show that for the second form (Eq.\ref{eq:bform2}) to yield good fit,
we must have $c_1<0$, which is physically unplausible given the
original motivation for this form of bias evolution. Thus, we choose
to use the first form (Eq.~\ref{eq:bform1}) to fit for biases
at different redshifts. The parameter $b_0$ and $\gamma$ are
then free parameters
in the MCMC with the priors given by $b_0\in(0,10)$ and $\gamma\in(-5,5)$.
The predicted values of bias
given by the best fit parameter for several samples (Main1, LRG, QSO3)
and those
given by \citet{wang08} are shown in Fig.~\ref{fig:b_f}. We see that the
functional form we adopt (Eq.\ref{eq:bform1}) is adequate to describe the
value of bias as a function of redshift for each sample.

Note that to see how the
LSS probes the PNG, here we do not consider joint constraint on
$f_{\rm NL}$ from CMB observations (e.g., bispectrum). We
also do not use the currently available CMB angular power spectrum data
in our MCMC unless otherwise stated (see next section about Planck prior).

\section{Results and Conclusions}
\label{sect:analysis}

\begin{figure}
\begin{center}
\includegraphics[width=0.6\textwidth]{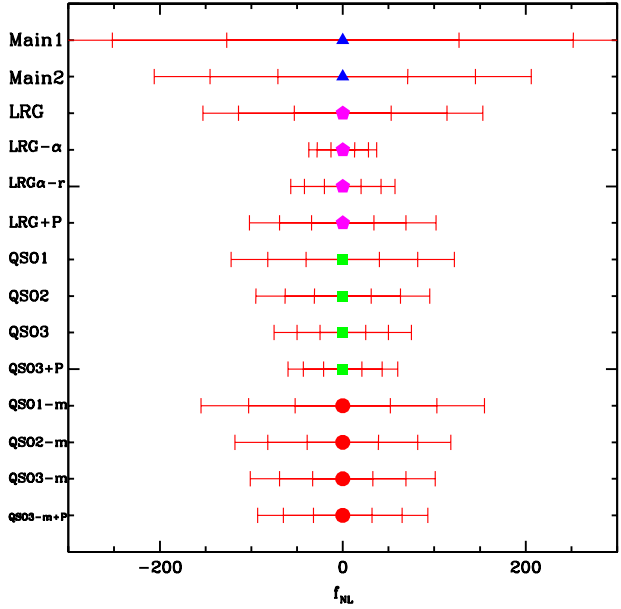}
\caption{\label{fig:fnl} The $1\sigma$, $2\sigma$, and $3\sigma$ range of
$f_{\rm NL}$ from LAMOST surveys. The samples are MAIN1 ($r<18.8$),
MAIN2 ($r<19.8$), LRG (the MegaZ sample), LRG-$\alpha$ (the constraint 
with the relative bias
method using the LRG and the photometric samples), LRG$\alpha$-r (the degraded
relative bias method considering bias evolution),
LRG+P (LRG with Planck priors),
QSO1 (QSO with $g<20.5$, non-merger case), QSO2 (QSO with $g<21$, non-merger case),
QSO3 (QSO with $g<21.65$, non-merger case), QSO3+P (QSO3 sample, non-merger case, 
with Planck priors),
QSO1-m (QSO1 sample, merger case), QSO2-m (QSO2 sample, merger case),
QSO3-m (QSO3 sample, merger case), QSO3-m+P (QSO3 sample, merger case, with Planck
priors). The assumed survey area is $8000\deg^2$.}
   \end{center}
\end{figure}

The result of our analysis, i.e., the $1\sigma(68.3\%),
2\sigma(95.5\%), 3\sigma(99.7\%)$ limits of $f_{\rm NL} $ inferred
from the power spectrum of the galaxy and quasar samples are given
in Table~\ref{tab:g} and Table~\ref{tab:q}, and also shown in
Fig.~\ref{fig:fnl}. We see that for the MAIN1 sample, the $3\sigma
(99.7\%)$ limit on $|f_{\rm NL}|$ reaches 379, so it provides only
very weak constraint on PNG. The MAIN2 and LRGs surveys have almost
the same effective volume on large scales \citep{wang08}, so they
provide comparable constraint on $f_{\rm NL}$. Comparing with the
current SDSS spectroscopic LRG survey \citep{slos08}, the constraint
on $|f_{\rm NL}|$ could be improved by a factor of about 1.5, to
reach a $2\sigma$ limit of about 114, which is comparable to current
limits from WMAP data \citep{koma08}.
%Note that here we assume
%that the area of the surveyed region for LAMOST is the same as SDSS.

Quasars are luminous and can be observed at high redshifts, so one
could obtain large survey volume with quasars. For the large scales
considered here, the quasars provide the largest effective volume
(c.f. Fig.3 of \citealt{wang08}), so we expect the LAMOST quasar
surveys to provide a better constraint on $f_{\rm NL}$. To account
for the possible effect of mergers, we calculated for the case of
$p=1$ (non-merger) and $p=1.6$ (merger). We see that even for the
QSO1 sample ($g<20.5$), the $95.5\% (2\sigma)$ limit on $|f_{\rm
NL}|$ reaches 103 for the merger case and 82 for the non-merger
case. For the QSO2 sample ($g<21$), the corresponding limits are
82(merger) and 63 (non-merger). For the QSO3 sample ($g<21.65$), the
limits shrink to 69 (merger) and 50 (non-merger). Comparing with the
SDSS quasar surveys \citep{slos08}, the LAMOST could improve the
constraint of $f_{\rm NL}$ by a factor of $1.3$. We note that we
assumed a lower QSO bias than \cite{slos08}, so the limits obtained
here are conservative.
%If it turns out that the bias of the
%QSO is higher, the constraint would be even stronger.

\begin{table}[h]
\begin{center}
\caption[]{\label{tab:g} The $1\sigma$, $2\sigma$, and $3\sigma$ range of
$f_{\rm NL}$ constrained with the LAMOST Galaxy Surveys. The last row
(LRG--$\alpha$) is from the relative bias method proposed by \citet{selj08}.}
%%Please Capitalize the First Letter of Each Notional Word in table's caption
  \begin{tabular}{crrr}
  \hline\hline\noalign{\smallskip}
Survey & $1\sigma$ & $2\sigma$ & $3\sigma$  \\
  \hline\noalign{\smallskip}
Main1 & $\pm 127$ & $\pm 252$ & $ \pm 379$ \\
Main2 & $\pm 71$ & $\pm 145$ & $\pm 206$  \\
LRG & $\pm 53$& $\pm 114$ & $\pm 153$  \\
LRG--$\alpha$ & $\pm 13$& $\pm 28$ & $\pm 37$  \\
  \noalign{\smallskip}\hline
  \end{tabular}
 \end{center}
\end{table}

\begin{table}[h]
\begin{center}
\caption[]{\label{tab:q} The $1\sigma$, $2\sigma$, and $3\sigma$ range of
$f_{\rm NL}$ constrained with the LAMOST Quasar Surveys.}
 \begin{tabular}{c|ccc|ccc}
  \hline\hline
Survey &  & Merger ($p=1.6$)&  &  & Non-merger ($p=1$)& \\
\cline{2-7}
 & $1\sigma$ & $2\sigma$ & $3\sigma$ & $1\sigma$ & $2\sigma$ & $3\sigma$ \\
  \hline
QSO1 & $\pm 52$ & $\pm 103$ & $\pm 155$ & $\pm 40$ & $\pm 82$ & $\pm 122$ \\
QSO2 & $\pm 39$ & $\pm 82$ & $\pm 118$ & $\pm 31$ & $\pm 63$ & $\pm 95$ \\
QSO3 & $\pm 33$& $\pm 69$ & $\pm 101$ & $\pm 25$ & $\pm 50$ & $\pm 75$ \\
\hline
  \end{tabular}\end{center}
\end{table}

The posterior probability distribution function
of $f_{\rm NL}$ appears to be nearly Gaussian here,
so the $1\sigma$, $2\sigma$, and $3\sigma$
errors are almost uniformly spaced. We have also investigated the
degeneracy between $f_{\rm NL}$ and $\Omega_{m0}$ and $n_s$.
We find that $f_{\rm NL}$ is positively correlated with $\Omega_{m0}$,
but it is almost independent with $n_s$, confirming the results
of \cite{slos08}.

By the time that the LAMOST surveys are completed, the Planck
satellite should have been taken CMB for a few years. With these CMB
data, the uncertainties on cosmological parameters should shrink
further. Here we simulate this effect by adding priors on the other
cosmological parameters based on the forecast of the Planck surveys.
We set the standard deviation as
$\Omega_{b0}=0.05\pm4.0\times10^{-4}$,
$\Omega_{m0}=0.24\pm5.0\times10^{-3}$,
$\sigma_8=0.82\pm6.7\times10^{-3}$, $n_s=0.96\pm4.0\times10^{-3}$
and $h_0=0.7\pm6.9\times10^{-3}$ \citep{colo08}. We present the
results for the LRG survey and QSO3 survey, and the impacts on other
cases are similar. With the stronger priors, the constraints are
improved slightly. With the QSO3 sample, we obtain $|f_{\rm NL}| <
65$ (merger case) and $|f_{\rm NL}| < 43$ (non-merger case) at
$95.5\%$ C.L., in comparison with $|f_{\rm NL}| < 69$ and $|f_{\rm
NL}| < 50$ without the Planck priors. These results are given in
Table 3 and also shown in Fig.\ref{fig:fnl}.

\begin{table}[h]
\begin{center}
\caption[]{\label{tab:Planck} The $1\sigma$, $2\sigma$, and $3\sigma$
constraints on $f_{\rm NL}$
from the LAMOST surveys with Planck priors.}
%%Please Capitalize the First Letter of Each Notional Word in table's caption
  \begin{tabular}{crrr}
  \hline\hline\noalign{\smallskip}
Survey & $1\sigma$ & $2\sigma$ & $3\sigma$  \\
  \hline\noalign{\smallskip}
LRG & $\pm 34$ & $\pm 69$ & $\pm 102$ \\
QSO3 merger & $\pm 32$ & $\pm 65$ & $\pm 93$  \\
QSO3 non-merger & $\pm 21$& $\pm 43$ & $\pm 60$  \\
  \noalign{\smallskip}\hline
  \end{tabular}
 \end{center}
\end{table}

%\begin{figure}
%   \begin{center}
%     \includegraphics[width=0.6\textwidth]{fnl_P.eps}
%   \caption{\label{fig:fnl_P}The $1,2,3\sigma$ error of
%$f_{NL}$ for LAMOST surveys with Planck prior.}
%   \end{center}
%\end{figure}

Comparing directly the density fields of tracers with different
biases in the same survey volume helps to eliminate the measurement error due
to cosmic
variance and break the degeneracy with other cosmological
parameters \citep{selj08, mcdo08b}. The relative bias
$\alpha=b_1/b_2$ is proposed to be used as the estimator of PNG
\citep{selj08,mcdo08b}. Under the assumption that the two tracers
have no stochasticity in them, and one tracer is unbiased (b=1), the
constraint of $f_{\rm NL}$ would be improved by a factor of
$\left(2\left[(\bar{n}_1P_1)^{-1}+(\bar{n}_2P_2)^{-1}\right]\right)^{-1/2}$.
The feasibility of this method depends on the number density of the
tracers. Here we consider the LRG as the high bias tracer, and a
sample of galaxies with photometric redshift as the low bias tracer.
The average redshift of the LRGs is $z \approx 0.4$. Using the halo
model method described in \cite{wang08}, we find that for a sample
of galaxies with good photometry up to $r<21.5$, the average
comoving density is $n=8.11 \times 10^{-3} h^3\Mpc^{-3}$, and the
average bias is $b=1.08$. This sample of galaxies satisfies the
requirement for the low bias tracer and has available SDSS
photometry. With the relative bias method, we find that the limit on
$f_{\rm NL}$ obtainable with LRGs could be improved by a factor of
about 4.1, with the $1\sigma$, $2\sigma$, and $3\sigma$ limits on
$|f_{\rm NL}|$ being 13, 28, and 37, respectively. 

In fact, since the evolution of the number density and clustering of the two
galaxy samples could be different,  the assumption of perfect 
correlation between the two tracer populations may not be valid,
especially for the large volumes used for the
non-Gaussinity test discussed here. This could introduce an additional 
error. The detailed study of this effect is beyond the scope of 
the present paper, we plan to investigate it in subsequent studies. 
For the problem discussed here, crude estimates show that in the 
worst case, it may cause a degration of the statistical power by as large 
as 50\%. The LRG$\alpha$-r in Fig.~\ref{fig:fnl} shows the degraded
precision.

In summary, we forecast the constraints on the primordial
non-Gaussianity from the LAMOST surveys. Using the scale-dependent
bias effect proposed by \cite{dala08}, and based on the estimations
of the galaxy and quasar surveys for the LAMOST \citep{wang08}, we
find that the galaxy surveys provide relatively weak constraint on
the $f_{\rm NL}$ parameter, but the QSO surveys could provide strong
constraints, with a $2\sigma$ limit on $|f_{\rm NL}|$ somewhere
between 50 to 103, depending on the sample size of the LAMOST QSO
survey. After adding the Plank priors for other cosmological
parameters, the QSO3 survey could give a $2\sigma$ limit of 43 for
$|f_{\rm NL}|$. We also considered using the relative bias method
\citep{selj08}. Correlating the LAMOST LRG sample with a photometric
sample of galaxies with $r<21.5$, for which $b\approx 1$ at $z=0.4$,
we obtain a $2\sigma$ limit of $|f_{\rm NL}|<28$. These results show
that the LAMOST could make useful constraints on the primordial
non-Gaussinity.

\begin{acknowledgements}

We thank An$\breve{\rm z}$e Slosar for discussions on details of
their calculation, and Pengjie Zhang for helpful suggestion. Our MCMC
chain computation was performed on the Supercomputing Center of the
Chinese Academy of Sciences and the Shanghai Supercomputing Center.
This work is supported by the National Science Foundation of China
under the Distinguished Young Scholar Grant 10525314, the Key
Project Grant 10533010, by the Chinese Academy of Sciences under the
grant KJCX3-SYW-N2, and by the Ministry of Science and Technology
national basic science Program (Project 973) under grant No.
2007CB815401. Z. Z. gratefully acknowledges support from the
  Institute for Advanced Study through a John Bahcall Fellowship.
\end{acknowledgements}

\label{lastpage}

\end{document}